\documentclass[12pt]{article}
\usepackage{amsmath,amssymb}
\usepackage{graphicx}
\usepackage{color}
\begin{document}
\baselineskip=16pt
\begin{titlepage}
\begin{flushright}
  {\small SU-HET-01-2015}
\\  {\small TU-987}
\end{flushright}
\vspace*{1.2cm}

\begin{center}

{\Large\bf Search for new physics via photon polarization of {\boldmath $b\rightarrow s\gamma$}}
\lineskip .75em
\vskip 1.5cm

\normalsize
{\large Naoyuki Haba}$^1$,
{\large Hiroyuki Ishida}$^1$, 
{\large Tsuyoshi Nakaya}$^2$,  

{\large Yasuhiro Shimizu}$^{3,4}$ and  
{\large Ryo Takahashi}$^1$

\vspace{1cm}
$^1${\it Graduate School of Science and Engineering, Shimane University, 

Matsue, 690-8504, Japan}\\

$^2${\it Department of Physics, Kyoto University, 

Kyoto, 606-8502, Japan}\\

$^3${\it Department of Physics, Tohoku University, 

  Sendai, 980-8578 Japan}
\\
$^4${\it Academic Support Center, Kogakuin University, 

  Hachioji,  192-0015 Japan}

\vspace*{10mm}

{\bf Abstract}\\[5mm]
{\parbox{13cm}{\hspace{5mm}
We suggest a discriminant analysis of new physics beyond the standard model 
through a detection of photon polarization in a radiative B meson decay. 
This analysis is investigated 
in SUSY SU(5) GUT with right-handed neutrino and left-right symmetric models.
New physics search via CP asymmetry in the same process are also evaluated 
in each model for comparison. 
We show that new physics can be found 
via detecting the photon polarization in a parameter space of TeV energy scale.
}}

\end{center}
\end{titlepage}

\section{Introduction}

In this decade, we have never seen an exotic elementary particle except for the 
most likely to the standard model (SM) Higgs boson which was discovered at the 
LHC experiment. This fact might imply that a scale of new physics (NP) is higher
 than the energy scale that the LHC can reach. Thus, it is important to consider
 the possibility that such exotics are hard to be directly produced by collider 
experiments even though the next LHC will become 14 TeV. 

On the other hand, indirect 
searches become powerful way to explore the existence of NP and related 
phenomena beyond the SM even if a new particle is impossible to be 
produced directly. One popular approach is flavor physics. For instance, flavor 
changing processes such as $\mu\rightarrow e\gamma$ and $b\rightarrow s\gamma$ 
retain much information about NP. Models of supersymmetric grand unified theory 
(SUSY GUT) predict relatively large branching ratios of Br$(\mu\rightarrow 
e\gamma)\sim\mathcal{O}(10^{-15} \mathchar`- 10^{-13})$ for the NP scale (the 
right-handed selectron mass) of $M_{\rm NP}\simeq(100 \mathchar`- 300)~{\rm GeV}$ 
in an SU(5) case and Br$(\mu\rightarrow e\gamma)\sim\mathcal{O}(10^{-13} 
\mathchar`- 10^{-11})$ in an SO(10) case~\cite{Barbieri:1994pv,Barbieri:1995tw}. 
Then, Ref.~\cite{Kuno:1996kv} suggested that the measurement of the angular 
distribution of $e$ with respect to the spin direction of the muon in the 
$\mu\rightarrow e\gamma$ process might distinguish among several extensions of 
the SM if the signal could be detected. This implies that the precise 
determination of the chirality of the final $e$ state in the $\mu\rightarrow 
e\gamma$ process might become a clue to obtain the evidence of NP. This 
situation is adopted to the $b\rightarrow s\gamma$ process, i.e. one would be 
able to discriminate among the SM and NP such as SUSY GUT models, the left-right
 symmetric standard model (LRSM), and the Pati-Salam models and so on, if one 
could precisely determine the chirality of the final $s$ quark. The $b$ quark 
can radiative decay into the $s$ quark in the $B$ meson and the chirality of
 the $s$ quark is almost determined as left-handed in the SM. Accordingly, if we
 find more right-handed $s$ quarks in the process than ones expected in the SM, 
we can expect that some kind of NP must cause this phenomena.

How about the measurement of the chirality of the $s$ quark in the $b\rightarrow
 s\gamma$ process for the NP search? One may naively think that the 
determination of chirality of quarks except for the top quark is impossible (the
top quark can decay before the hadronization). 
The $b\rightarrow s\gamma$ process occurs through the dipole type operators,
 $\overline{s}_{L}\sigma_{\mu\nu}bF^{\mu\nu}$ 
( $\overline{s}_{R}\sigma_{\mu\nu}bF^{\mu\nu}$) which induce left- ( right- ) handed photon. The information on the chirality of the $s$ quark is imprinted on the photon polarization.
In addition, there is no parity violation in QCD, the relation between the chirality of the $s$ quark and the photon polarization is unchanged even if the hadronization is taken into account. Therefore, one can determine the 
chirality of $s$ quark in the $b\rightarrow s\gamma$ process from the 
measurement of photon polarization \cite{Gronau:2001ng,Gronau:2002rz}.
In Refs. \cite{Grinstein:2004uu,Keum:2004is}, 
the authors mentioned that the higher order correction 
may induce the right-handed photon even though it is in the SM.

At $e^+e^-$ colliders, such as Belle and BaBar, $B_d$ mesons, which are spin 0 particles, are produced from the $\Upsilon(4S)$ resonance. 
The photon polarization of the $B_d \rightarrow X_s\gamma$ decay is  
determined from measurements of hadronic angular distributions 
due to the conservation of angular momentum.
The LHCb collaboration actually reported the result of observation of photon polarization by measuring the angular distribution of produced mesons in the $B\rightarrow K\pi\pi\gamma$ 
process~\cite{Aaij:2014wgo}.\footnote{This issue tells us that the theorists 
have to clarify the prediction of photon polarization in each model.} In 
addition to the $B\rightarrow 
K\pi\pi\gamma$ process, there is another possibility to determine the photon polarization by the $B\rightarrow 
K^\ast l^+l^-$ process. Although there exists a box diagram in the process, the 
radiative decay diagram (penguin) becomes dominant (the box diagram is suppressed) in a low 
 invariant mass region of dileptons~\cite{Melikhov:1998cd,Kim:2000dq} (see 
also~\cite{Becirevic:2012dx}).
The chirality of the $s$ quark in the $K^*$ meson can be lead by the chirality of photon 
due to the conservation of the spin. 
Then, it is important to discuss the possibility of the detection of the photon chirality.
In this work, we will consider the $b\rightarrow s\gamma$ process. In 
particular, a ratio of the Wilson coefficients of a dipole operator and a 
polarization parameter of photon will be firstly evaluated at a typical point in a model 
of SUSY SU(5) GUT with the right-handed neutrino ($N_R$) and LRSM in order to clarify 
whether one can find an evidence of NP or distinguish among the SM and NP, or not.

In addition to the determination of  the photon chirality, the CP asymmetry
 in the $b\rightarrow s\gamma$ process 
which are direct CP asymmetry, $A_{\rm CP}(b\rightarrow s\gamma)$ 
and time-dependent CP asymmetry, $S_{\rm CP}(B\rightarrow K_s \pi^0 \gamma)$, is 
also a sensitive observable to NP~\cite{Soares:1991te}. Actually, the CP 
violating effects from NP can be enough larger than the SM expectation as 
$A_{\rm CP}(b\rightarrow s\gamma)\simeq-0.5\%$. 
However, this CP violation is 
constrained by the other experiment, e.g. the chromo electric dipole moment 
(CEDM)~\cite{Endo:2010yt}, and it can be negligibly smaller than the SM one when
 the CP violating phase depending on the $b\rightarrow s\gamma$ process is 
accidentally small. We will also evaluate the magnitude of $A_{\rm 
CP}(b\rightarrow s\gamma)$ in the SUSY SU(5) GUT with $N_R$ 
and the LRSM although the magnitude strongly depends on CP violating phases in the models. 
Furthermore, $S_{\rm CP}(B\rightarrow K_s \pi^0 \gamma)$ can also become 
larger than the SM expectation as $S_{\rm CP}(B\rightarrow K_s \pi^0 \gamma) \simeq -0.3$.
We will show this value is insensitive to the CP phase. 
Then, we will compare experimental detectability for the models of NP between the 
determination of photon polarization and the observation of both CP asymmetry in the 
$b\rightarrow s\gamma$ process.
 
We will suggest that one can discriminate NP beyond the SM by the 
detection of photon polarization in $b\rightarrow s \gamma$ process. 
We will point out that time-dependent CP asymmetry is the 
most stringent constraint in our sample model point at the moment.
However, it will actually turn out that the LHCb with 
2 fb$^{-1}$ for the determination of photon polarization may check the existence 
of NP scale up to several TeV in both models.

\section{Photon polarization}
We investigate the photon polarization in the radiative rare decay, 
$b\rightarrow s\gamma$, process in a SUSY SU(5) GUT with $N_R$ 
and the LRSM for the search of NP. The Wilson coefficients $C_7$ and 
$C_7'$ of the dipole operator for the $b\rightarrow s\gamma$ process 
are important for the analyses of photon polarization.
The effective Hamiltonian
 reads
 \begin{eqnarray}
  \mathcal{H}_{\rm eff}\supset
  -\frac{4G_F}{\sqrt{2}}V_{tb}V_{ts}^\ast(C_7O_7+C_7'O_7'),
 \end{eqnarray}
with the magnetic operator,
 \begin{eqnarray}
  O_7=\frac{e}{16\pi^2}m_b(\bar{s}\sigma^{\mu\nu}P_Rb)F_{\mu\nu},
 \end{eqnarray}
where $G_F$ is the Fermi constant, $m_b$ is the bottom quark mass, 
$\sigma^{\mu\nu}=\frac{i}{2}[\gamma^\mu,\gamma^\nu]$, and 
$P_{R,L}=\frac{1}{2}(1\pm\gamma_5)$~(e.g., see~\cite{Buras:1998raa}). $O_7'$ is 
obtained by replacing $L\leftrightarrow R$ in $O_7$. 
Because left-handed $s$ quark comes from $O_7$ and right-handed one comes from $O'_7$, we might be able to determine the chirality of $s$ quark by the difference of Wilson coefficients.

When one considers physics 
beyond the SM, there might be additional contributions to $C_7$ and $C_7'$ from 
NP. In these cases, we can generically describe $C_7$ and $C_7'$ as $C_7=C_7^{\rm 
SM}+C_7^{\rm NP}$ and $C_7'=C_7'{}^{\rm SM}+C_7'{}^{\rm NP}$, respectively. The 
coefficients at the $b$ quark mass scale $\mu_b$ are given by the leading 
logarithmic calculations with QCD corrections to the $b\rightarrow s\gamma$ 
process, 
 \begin{eqnarray}
  C_7(\mu_b)  &=& \eta^{\frac{16}{23}}C_7(m_W)
                  +\frac{8}{3}(\eta^{\frac{14}{23}}-\eta^{\frac{16}{23}})C_8(m_W)
                  +\sum_{i=1}^8h_i\eta^{a_i}, \\
  C_7'(\mu_b) &=& \eta^{\frac{16}{23}}C_7'(m_W)
                  +\frac{8}{3}(\eta^{\frac{14}{23}}-\eta^{\frac{16}{23}})C_8'(m_W),
 \end{eqnarray}
at the leading order where $\eta=\alpha_s(m_W)/\alpha_s(\mu_b)$, $\alpha_s\equiv 
g_s^2/(4\pi)$, $g_s$ is the strong coupling constant, $m_W$ is the $W$ boson 
mass, and $h_i$ and $a_i$ are numerical 
coefficients~\cite{Kagan:1998ym,Ciuchini:1993ks,Buras:1993xp}.  $C_8$ is the 
coefficient of chromomagnetic operator
 \begin{eqnarray}
  O_8=\frac{g_s}{16\pi^2}m_b(\bar{s}\sigma^{\mu\nu}T^AP_Rb)G_{\mu\nu}^A,
 \end{eqnarray}
in the $\Delta F=1$ effective Hamiltonian,
 \begin{eqnarray}
  \mathcal{H}_{\rm eff}\supset
  -\frac{4G_F}{\sqrt{2}}V_{tb}V_{ts}^\ast(C_8O_8+C_8'O_8'),
 \end{eqnarray}
where $T^A$ are the generators of $SU(3)_C$ and $O_8'$ is also obtained by 
replacing $L\leftrightarrow R$ in $O_8$. And, we also describe 
$C_8=C_8^{\rm SM}+C_8^{\rm NP}$ and $C_8'=C_8'{}^{\rm SM}+C_8'{}^{\rm NP}$ including 
contributions from NP.

\subsection{Case of SUSY SU(5) with the right-handed neutrinos}

\subsubsection{Model}

At first, we give a brief review of a model of SUSY SU(5) GUT with $N_R$.
In a simple SU(5) GUT model, 
the final $s$ quark must have the same chirality as in the SM.
When there is $N_R$, a neutrino Yukawa coupling induces additional flavor mixings in the 
right-handed down squark which derives the opposite chirality of $s$ quark. 
Thus, we adopt the model of SUSY SU(5) GUT with $N_R$.
The superpotential in this model is given by
 \begin{eqnarray}
  W=\frac{1}{4}f_{ij}^u\Psi_i\Psi_jH+\sqrt{2}f_{ij}^d\Psi_i\Phi_j\bar{H}
    +f_{ij}^\nu\Phi_i\bar{N}_jH+M_{ij}\bar{N}_i\bar{N}_j,
 \end{eqnarray}
where $\Psi_i$ are {\bf 10}-dimensional multiplets, $\Phi_i$ are 
{\bf 5}-dimensional ones, $N_i$ denote the right-handed neutrino superfields, 
and $H$ ($\bar{H}$) is {\bf 5}- ($\bar{\mbox{{\bf 5}}}$-) dimensional Higgs 
multiplets. $i$ and $j$ mean the generation of the fermions, $i,j=1,2,3$. $f^u$,
 $f^d$, and $f^\nu$ are Yukawa coupling matrices for the up-type quarks, 
down-type quarks (charged leptons) and neutrinos, respectively. These are given 
by
 \begin{eqnarray}
  f_{ij}^u &=& V_{ki}f_{u_k}e^{i\varphi_{u_k}}V_{kj}, \\
  f_{ij}^d &=& f_{d_i}\delta_{ij}, \\
  f_{ij}^\nu &=& e^{i\varphi_{d_i}}U_{ij}^\ast f_{\nu_j},
 \end{eqnarray}
without a loss of generality, where $V$ and $U$ are the 
Cabibbo-Kobayashi-Maskawa (CKM) and Pontecorvo-Maki-Nakagawa-Sakata (PMNS) 
matrices, respectively. $\varphi_{u_k}$ and $\varphi_{d_i}$ are CP-violating 
phases, and $f_{u_k}$ and $f_{d_i}$ are Yukawa couplings of the up- and 
down-type quarks (charged leptons), respectively. For the neutrinos sector, the 
light neutrino masses are given by the seesaw mechanism 
$m_{\nu_i}=f_{\nu_i}^2v_u^2/M_{N_i}$, where $v_u$ is the vacuum expectation value 
(VEV) of the up-type Higgs in $H$, and $M_{N_i}$ are the mass eigenvalues of the 
right-handed neutrinos. Here, we assume a diagonal right-handed Majorana mass 
matrix $M_{ij}$ for simplicity.

\subsubsection{Photon polarization in SUSY SU(5) with right-handed neutrino}

We discuss $C_7$ and $C_7'$, which determine the magnitude of the photon 
polarization, in the model of SUSY SU(5) GUT with $N_R$. In
 supersymmetric models, the dominant contributions to $C_{7,8}$ and $C_{7,8}'$ 
arise from loop diagrams of the charged Higgses, charginos, and gluinos. Thus, 
$C_{7,8}^{\rm NP}=C_{7,8}^{H^\pm}+C_{7,8}^{\tilde{\chi}^\pm}+C_{7,8}^{\tilde{g}}$ and 
$C_{7,8}'^{\rm NP}=C_{7,8}'^{H^\pm}+C_{7,8}'^{\tilde{\chi}^\pm}+C_{7,8}'^{\tilde{g}}$ in
 the SUSY SU(5) model with $N_R$ where $C_{7,8}^{H^\pm}$ ($C_{7,8}'^{H^\pm}$), 
$C_{7,8}^{\tilde{\chi}^\pm}$ ($C_{7,8}'^{\tilde{\chi}^\pm}$), and $C_{7,8}^{\tilde{g}}$ 
($C_{7,8}'^{\tilde{g}}$) are the contributions to $C_{7,8}$ ($C_{7,8}'$) from the 
charged Higgses, charginos, and gluinos, respectively. These contributions are 
calculated as~\cite{Altmannshofer:2009ne}
 \begin{eqnarray}
  C_7^{H^\pm} &=& C_7'{}^{H^\pm}\simeq
   \left(\frac{1-\epsilon t_\beta}{1+\epsilon t_\beta}\right)\frac{1}{2}h_7(y_t), 
   \\
  C_7^{\tilde{\chi}^\pm} 
   &=& \frac{4G_F}{\sqrt{2}}\frac{g_2^2}{\tilde{m}^2}
       \left[\frac{(\delta_u^{LL})_{32}}{V_{tb}V_{ts}^\ast}
             \frac{\mu M_2}{\tilde{m}^2}f_7^{(1)}(x_2,x_\mu)
             +\frac{m_t^2}{M_W^2}\frac{A_t\mu}{\tilde{m}^2}f_7^{(2)}(x_\mu)\right]
       \frac{t_\beta}{1+\epsilon t_\beta}, \\
  C_7'{}^{\tilde{\chi}^\pm} 
   &=& \frac{4G_F}{\sqrt{2}}\frac{g_2^2}{\tilde{m}^2}
       \left[\frac{(\delta_u^{RR})_{32}}{V_{tb}V_{ts}^\ast}
             \frac{\mu M_2}{\tilde{m}^2}f_7^{(1)}(x_2,x_\mu)
             +\frac{m_t^2}{M_W^2}\frac{A_t\mu}{\tilde{m}^2}f_7^{(2)}(x_\mu)\right]
       \frac{t_\beta}{1+\epsilon t_\beta}, \\
  C_7^{\tilde{g}^\pm} 
   &=& \frac{4G_F}{\sqrt{2}}\frac{g_s^2}{\tilde{m}^2}
       \left[\frac{M_{\tilde{g}}}{m_b}\frac{(\delta_d^{RL})_{32}}{V_{tb}V_{ts}^\ast}
             g_7^{(1)}(x_g)
             +\frac{M_{\tilde{g}}\mu}{\tilde{m}^2}\frac{t_\beta}{1+\epsilon t_\beta}
              \frac{(\delta_d^{LL})_{32}}{V_{tb}V_{ts}^\ast}g_7^{(2)}(x_g)\right], \\
  C_7'{}^{\tilde{g}^\pm} 
   &=& \frac{4G_F}{\sqrt{2}}\frac{g_s^2}{\tilde{m}^2}
       \left[\frac{M_{\tilde{g}}}{m_b}\frac{(\delta_d^{LR})_{32}}{V_{tb}V_{ts}^\ast}
             g_7^{(2)}(x_g)
             +\frac{M_{\tilde{g}}\mu^\ast}{\tilde{m}^2}
              \frac{t_\beta}{1+\epsilon t_\beta}
              \frac{(\delta_d^{RR})_{32}}{V_{tb}V_{ts}^\ast}g_7^{(1)}(x_g)\right],
 \end{eqnarray}
and 
  \begin{eqnarray}
  C_8^{H^\pm} &=& C_8'{}^{H^\pm}\simeq
   \left(\frac{1-\epsilon t_\beta}{1+\epsilon t_\beta}\right)\frac{1}{2}h_8(y_t), 
   \\
  C_8^{\tilde{\chi}^\pm} 
   &=& \frac{4G_F}{\sqrt{2}}\frac{g_2^2}{\tilde{m}^2}
       \left[\frac{(\delta_u^{LL})_{32}}{V_{tb}V_{ts}^\ast}
             \frac{\mu M_2}{\tilde{m}^2}f_8^{(1)}(x_2,x_\mu)
             +\frac{m_t^2}{M_W^2}\frac{A_t\mu}{\tilde{m}^2}f_8^{(2)}(x_\mu)\right]
       \frac{t_\beta}{1+\epsilon t_\beta}, \\
  C_8'{}^{\tilde{\chi}^\pm} 
   &=& \frac{4G_F}{\sqrt{2}}\frac{g_2^2}{\tilde{m}^2}
       \left[\frac{(\delta_u^{RR})_{32}}{V_{tb}V_{ts}^\ast}
             \frac{\mu M_2}{\tilde{m}^2}f_8^{(1)}(x_2,x_\mu)
             +\frac{m_t^2}{M_W^2}\frac{A_t\mu}{\tilde{m}^2}f_8^{(2)}(x_\mu)\right]
       \frac{t_\beta}{1+\epsilon t_\beta}, \\
  C_8^{\tilde{g}^\pm} 
   &=& \frac{4G_F}{\sqrt{2}}\frac{g_s^2}{\tilde{m}^2}
       \left[\frac{M_{\tilde{g}}}{m_b}\frac{(\delta_d^{RL})_{32}}{V_{tb}V_{ts}^\ast}
             g_8^{(1)}(x_g)
             +\frac{M_{\tilde{g}}\mu}{\tilde{m}^2}\frac{t_\beta}{1+\epsilon t_\beta}
              \frac{(\delta_d^{LL})_{32}}{V_{tb}V_{ts}^\ast}g_8^{(2)}(x_g)\right], \\
  C_8'{}^{\tilde{g}^\pm} 
   &=& \frac{4G_F}{\sqrt{2}}\frac{g_s^2}{\tilde{m}^2}
       \left[\frac{M_{\tilde{g}}}{m_b}\frac{(\delta_d^{LR})_{32}}{V_{tb}V_{ts}^\ast}
             g_8^{(2)}(x_g)
             +\frac{M_{\tilde{g}}\mu^\ast}{\tilde{m}^2}
              \frac{t_\beta}{1+\epsilon t_\beta}
              \frac{(\delta_d^{RR})_{32}}{V_{tb}V_{ts}^\ast}g_8^{(1)}(x_g)\right],  
 \end{eqnarray}
at the weak  scale and 
$\epsilon\simeq\alpha_s/(3\pi)\sim\mathcal{O}(10^{-2})$ for a degenerate SUSY 
spectrum, $t_\beta=\tan\beta\equiv v_u/v_d$, $v_d$ is the VEV of down-type Higgs, 
$g_2$ is the $SU(2)_L$ gauge coupling constant, $\tilde{m}$ is an averaged squark
 mass, $(\delta_q^{XY})_{ij}$ ($q=u,d$ and $X,Y=L,R$) are mass insertion 
parameters, $\mu$ is the supersymmetric Higgs mass, $M_x$ ($x=2,\tilde{g}$) are 
the gaugino masses, and $A_t$ is the soft scalars coupling for the top quark. 
$h_{7,8}$, $f_{7,8}^{(1,2)}$, and $g_{7,8}^{(1,2)}$ are loop functions, which are given
 in Appendix A. The mass insertion parameters are given in Appendix B. And, we 
define $y_t\equiv m_t^2/M_{H^\pm}^2$, $x_2\equiv|M_2|^2/\tilde{m}^2$, 
$x_\mu\equiv|\mu|^2/\tilde{m}^2$, and $x_g\equiv M_{\tilde{g}}^2/\tilde{m}^2$ where 
$m_t$ and $M_{H^\pm}$ are the top quark and charged Higgs masses, respectively. 
The contributions from the charged Higgs and chargino to $C_7'{}^{\rm NP}$ are 
suppressed by $m_s/m_b$.

\begin{figure}
\begin{center}
\includegraphics[scale=0.75]{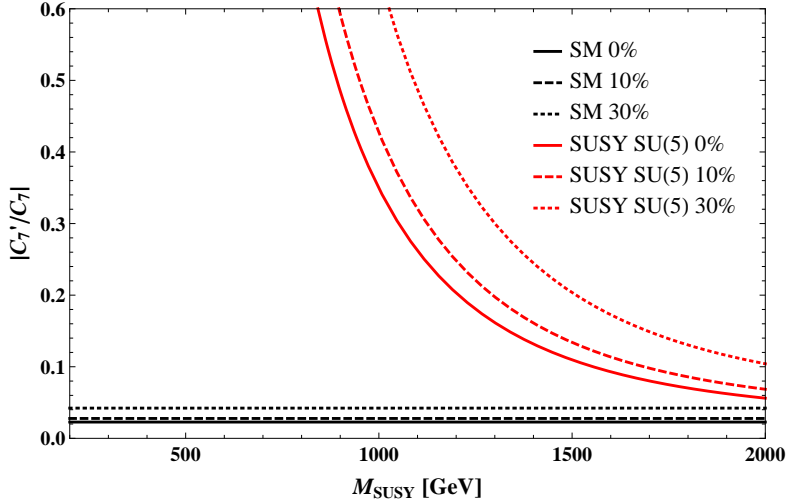}
\end{center}
\caption{The magnitude of $|C_7'/C_7|$ in the SM and SUSY SU(5) GUT with $N_R$ 
which are depicted by the black and red curves, 
respectively. The contours with $n\%$ correspond to cases that there is $n\%$ 
misidentification in the $C_7$ measurement at experiments, i.e. each of the 
contours denotes $|C_7'(1+n/100)/(C_7(1-n/100))|$.}
\label{fig1}
\end{figure}

In order to see the magnitude of contributions from NP, we estimate the ratio 
$|C_7'/C_7|$ at the $b$ quark mass scale, which determines the size of 
polarization of photon as seen below. The value of the ratio in the model of 
SUSY SU(5) GUT with $N_R$ is shown by the red solid curve in 
Fig.~\ref{fig1}. 
The black solid line indicates the SM case $|C_7'{}^{\rm SM}/C_7^{\rm SM}|$ which 
can be approximated as $|C_7'{}^{\rm SM}/C_7^{\rm SM}|\simeq m_s/m_b$. One can see 
that $|C_7'/C_7|$ in the SUSY SU(5) with $N_R$ case is enhanced from the SM. This means 
that the final state of $s_R$ in $b\rightarrow s\gamma$ increases compared to 
the SM while the most of final state of $b\rightarrow s\gamma$ in the SM is 
$s_L$ due to the suppression proportional to $m_s/m_b$. 

In Fig.~\ref{fig1}, the horizontal axis is a typical scale of NP, which is a 
SUSY breaking scale $M_{\rm SUSY}$ in the SUSY SU(5) GUT with $N_R$ case. 
One can see that at
 a large limit of $M_{\rm SUSY}$, the ratio $|C_7'/C_7|$ closes to the SM case, 
$|C_7'/C_7|\rightarrow |C_7'{}^{\rm SM}/C_7^{\rm SM}|$. The contours with $n$\% 
correspond to case that there is $n$\% misidentification in the $C_7$ 
measurement at experiments, i.e. the contours denote 
$|C_7'(1+n/100)/(C_7(1-n/100))|$. This misidentification corresponds to a 
mismatch in the conversion of left-handed helicity to the left-handed chirality.
 (The helicity is determined in experiments.) For instance, if one identifies 
the left-handed helicity with the left-handed chirality with $10\%$ 
misidentification, the right-handed chirality is over estimated as $110\%$ of 
the true value. Thus, the contours go above as $n$ increases in Fig.~\ref{fig1}.
 In the calculation, we take
 \begin{eqnarray}
  && t_\beta=10,~~~M_{\rm SUSY}=m_{1/2},~~~m_0=A_0=A_t=A_b=1~{\rm TeV},~~~
     \mu=1.01m_0, \\
  && M_2=0.822M_{\rm SUSY},~~~M_{\tilde{g}}=2.86 M_{\rm SUSY},~~~
     M_{H_c}=10^{16}~{\rm GeV},~~~
     \tilde{m}^2_{\tilde{u}}=\sqrt{m_{\tilde{Q}_L}^2m_{\tilde{u}_R}^2},~~~\\
  && \tilde{m}^2_{\tilde{d}}=\sqrt{m_{\tilde{Q}_L}^2m_{\tilde{d}_R}^2},~~~
     m_{\tilde{Q}_L}^2=m_0^2+6.86M_{\rm SUSY}^2,~~~
     m_{\tilde{u}_R}^2=m_0^2+6.44M_{\rm SUSY}^2,~~~\\
  && m_{\tilde{d}_R}^2=m_0^2+6.39M_{\rm SUSY}^2,~~~
     f_{\nu_i}=1,~~~ \varphi_{u_{23}}=0.01,~~~\varphi_{d_{23}}=\pi, 
 \end{eqnarray}
as a typical point where CP phases are given in the radian 
unit.\footnote{Note that $\varphi_{d_{23}}$ is sensitive to the CEDM. The allowed 
minimal and maximal values by the CEDM constraint are $\pi$ and $3\pi/2$ (or 
$\pi/2$), respectively. In the calculation of photon polarization, we take the 
minimal value. The photon polarization is not so sensitive to the value of 
$\varphi_{d_{23}}$ but the difference between these values appears in the 
calculation of direct CP asymmetry as we will show in Section 3. We have 
numerically checked that another phase, $\varphi_{u_{23}}$, is not sensitive to 
our evaluation.} We take other values of parameters in the SM and the neutrino 
sector (PMNS mixing angles) as the best fit values given 
in~\cite{pdg,Capozzi:2013csa}.

Next, we consider the polarization parameter of photon $\lambda_\gamma$ at the $b$ 
quark mass scale defined as
 \begin{eqnarray}
  \lambda_\gamma\equiv\frac{{\rm Re}[C_7'/C_7]^2+{\rm Im}[C_7'/C_7]^2-1}
                          {{\rm Re}[C_7'/C_7]^2+{\rm Im}[C_7'/C_7]^2+1}.
 \end{eqnarray}
In order to measure $\lambda_\gamma$ we need to consider a parity-odd 
observable in the $B_d\to X_s\gamma$ decay since the photon polarization is 
parity-odd.  In Ref.\cite{Gronau:2001ng,Gronau:2002rz} they proposed that 
$\lambda_\gamma$ can be measured from the up-down asymmetry of the photon 
direction relative to the $K\pi\pi$ decay plane in the $K_1 (1400)$ rest frame.
\begin{figure}
\begin{center}
\includegraphics[scale=0.82]{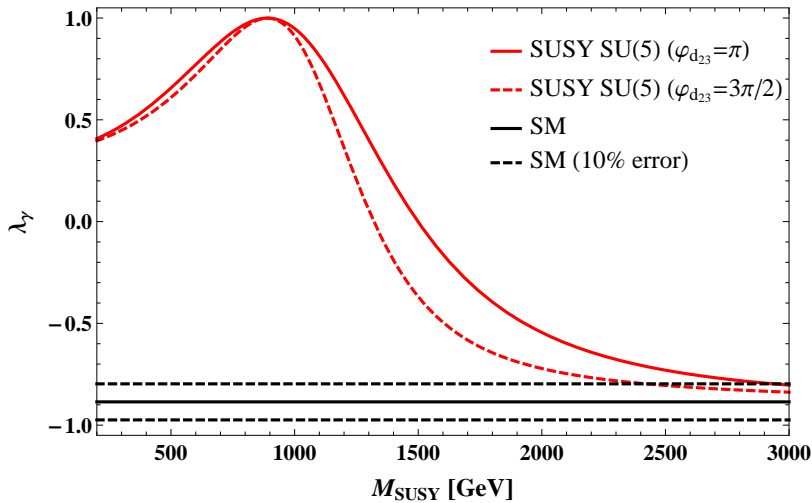}
\end{center}
\caption{The polarization parameter $\lambda_\gamma$ in $b\rightarrow s\gamma$ in
 the SM and SUSY SU(5) GUT with $N_R$ which are 
depicted by the black and red curves, respectively. We also show 
$\varphi_{d_{23}}=3\pi/2$ case by the red dashed curve 
and $10\%$ error in the SM prediction by the black dashed curve} for comparison.
\label{fig2}
\end{figure}
We show $\lambda_\gamma$ in Fig.~\ref{fig2} and 
the SM and SUSY SU(5) GUT with $N_R$ cases are depicted by 
the black and red curves, respectively. We also show $\varphi_{d_{23}}=3\pi/2$ 
case by the red dashed curve for comparison. 
The SM prediction for the photon polarization still have roughly $10\%$ error due to $\mathcal{O}_2$ operator~\cite{Becirevic:2012dx}, 
then we depict it by the black dashed curve.

The superKEKB  with an
integrated luminosity of $75~{\rm ab}^{-1}$ and the LHCb with $2~{\rm fb}^{-1}$ 
might reach the 20 ($|\lambda_\gamma|=0.8 |\lambda_\gamma^{\rm SM}|$) and $10\%$ ($|\lambda_\gamma|=0.9 |\lambda_\gamma^{\rm SM}|$) 
precision, respectively~\cite{Becirevic:2012dx,Kou:2010kn}. Thus, the future 
experiment will be able to check the NP scale up to about 1700 GeV (which 
corresponds to $M_2\simeq1400$ GeV and $M_{\tilde{g}}\simeq4900$ GeV) in this model 
of SUSY SU(5) with $N_R$.

\subsection{Case of Left-right symmetric standard model}

\subsubsection{Model}

Next, we consider the case in the left-right symmetric standard model 
(LRSM)~\cite{Pati:1974yy,Mohapatra:1974hk,Mohapatra:1974gc,Senjanovic:1975rk}. 
The model is based on the gauge group $SU(2)_L\times SU(2)_R\times 
U(1)_{\tilde{Y}}$. In the model, the SM left-handed doublet fermions are $SU(2)_R$ 
singlets, and the right-handed fermions including the neutrinos are $SU(2)_R$ 
doublets and $SU(2)_L$ singlets. For the Higgs sector, the model includes a 
bi-doublet scalar $\Phi$ under the $SU(2)_L\times SU(2)_R$ transformation, an $SU(2)_R$ triplet 
$\Delta_R$, and an $SU(2)_L$ triplet $\Delta_L$ in order to realize a realistic 
symmetry breaking, $SU(2)_L\times SU(2)_R\times U(1)_{\tilde{Y}}\rightarrow 
SU(2)_L\times U(1)_Y\rightarrow U(1)_{\rm em}$. The symmetry breaking can be 
undertaken by the VEVs of $\Phi$, $\Delta_R$, and $\Delta_L$ as
 \begin{eqnarray}
  \langle\Phi\rangle=\left(
                      \begin{array}{cc}
                       \kappa & 0 \\
                       0 & \kappa'e^{i\omega}
                      \end{array}
                     \right),~~~
  \langle\Delta_R\rangle=\left(
                         \begin{array}{cc}
                          0  & 0 \\
                          v_R & 0
                         \end{array}
                        \right),~~~ 
  \langle\Delta_L\rangle=\left(
                         \begin{array}{cc}
                          0            & 0 \\
                          v_Le^{i\theta_L} & 0
                         \end{array}
                        \right),
 \end{eqnarray}
with six real numbers $\kappa$, $\kappa'$, $\omega$, $v_R$, $v_L$, and 
$\theta_L$. Regarding the magnitude of the VEVs, $v_R$ should be much larger 
than the electroweak (EW) scale  to suppress the right-handed currents at low 
energy, and the EW $\rho$-parameter limits $v_L$ to be $v_L\lesssim10$ 
GeV~\cite{Amaldi:1987fu}. In this work, we take $v_L=0$ for simplicity, which is
 usually taken in literatures (e.g., \cite{Blanke:2011ry,Kou:2013gna}). Since 
the VEV of $\Phi$ leads to the standard EW symmetry breaking, we define 
$v\equiv\sqrt{\kappa^2+\kappa'{}^2}=174$ GeV, $\tan\beta_{\rm 
LR}\equiv\kappa/\kappa'$, and $\epsilon_{\rm LR}\equiv v/v_R$.

The charged gauge bosons are given by the admixture of the mass eigenstates as 
 \begin{eqnarray}
  \begin{pmatrix}
	W_L^-\\
	W_R^-
  \end{pmatrix}
	&=&
  \begin{pmatrix}
	\cos \zeta &- \sin \zeta e^{i \omega}\\
	\sin \zeta e^{i \omega} &\cos \zeta
  \end{pmatrix}
  \begin{pmatrix}
	W_1^-\\
	W_2^-
  \end{pmatrix}.
 \end{eqnarray}
The masses of charged gauge bosons are approximated as 
 \begin{eqnarray}
  M_{W_1}\simeq\frac{g_Lv}{\sqrt{2}}
              (1-\epsilon_{\rm LR}^2\sin^2\beta_{\rm LR}\cos^2\beta_{\rm LR}),~~~
  M_{W_2}\simeq g_Rv_R(1+\frac{1}{4}\epsilon_{\rm LR}^2),
 \end{eqnarray}
where $g_{L,R}$ are the gauge couplings of $SU(2)_{L,R}$ and we take $g_R/g_L=1$ 
for simplicity in the numerical analysis. The mixing angle is given as 
$\sin \zeta \approx (M_{W_1}^2/M_{W_2}^2) \sin 2 \beta_{LR}$.
$M_{W_2}$ is identified with $M_{\rm NP}$ in the LRSM.
 There are also charged and heavy neutral Higgs bosons in the LRSM. And their 
masses are nearly the same,
  $M_{H^\pm}\simeq M_{H^0}\simeq M_{A^0}$,
where $M_{H^\pm}$ are the masses of the charged Higgs bosons and $M_{H^0,A^0}$ are 
the neutral Higgs bosons masses~\cite{Blanke:2011ry,Zhang:2007da}. In this work,
 we represent both the charged and neutral Higgs bosons masses as $M_H$ for 
simplicity. Regarding the flavor mixing matrices, we assume $V=V_L=V_R$, where 
$V_L$ and $V_R$ are the mixing matrices for the left- and right-handed quarks, 
respectively. $V=V_L=V_R$ is taken in the so-called the manifest 
LRSM~\cite{Senjanovic:1978ev,Beg:1977ti} and we also take the equality in this 
work. In our numerical analyses, we have three free parameters, i.e. $M_H$, 
$M_{\rm NP}=M_{W_2}$, and $\tan\beta_{\rm LR}$.\footnote{The value of $M_{W_2}$ is not
 exactly determined even if one takes $g_R/g_L=1$ and fixes the value of $M_H$, 
because the heavy Higgs masses depend on scalar quartic couplings, which can be 
in region from 0 to $4\pi$, and/or trilinear couplings. Thus, one can generally 
take both $M_{W_2}$ and $M_H$ as free parameters in this model.}

\subsubsection{Photon polarization in LRSM}

\begin{figure}
\begin{center}
\includegraphics[scale=0.78]{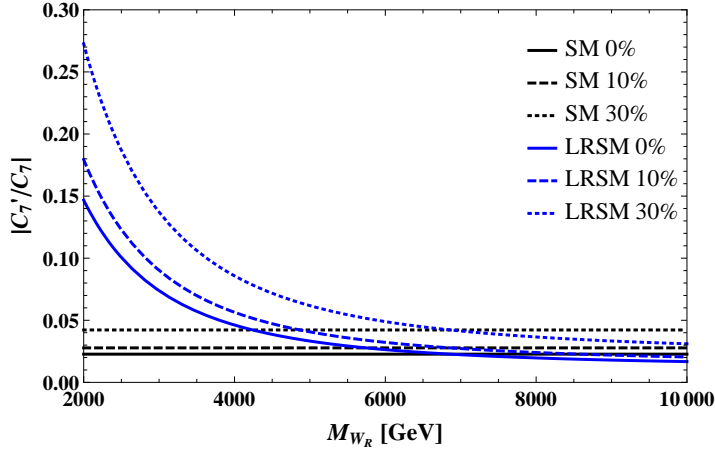}
\end{center}
\caption{The magnitude of $|C_7'/C_7|$ in the SM and LRSM which are depicted by 
the black and blue curves, respectively. The meaning of each contour is the same
 as Fig. \ref{fig1}.}
\label{fig3}
\end{figure}

$C_7^{\rm NP}$ and $C_7'{}^{\rm NP}$ in the LRSM are generally given by
 \begin{eqnarray}
  C_7^{\rm NP} &=& -\sin^2\zeta \left(D_0'(x_t)
                -\frac{M_{W_1}^2}{M_{W_2}^2}D_0'(\tilde{x}_t)\right) \nonumber \\
             & & +\frac{m_t}{m_b}\frac{g_R}{g_L}\frac{V_{tb}^R}{V_{tb}}\sin\zeta
                 \cos\zeta e^{i\omega}\left(A_{\rm LR}(x_t)-
                 \frac{M_{W_1}^2}{M_{W_2}^2}A_{\rm LR}(\tilde{x}_t)\right) \nonumber 
                 \\
             & & +\frac{m_c}{m_b}\frac{g_R}{g_L}
                  \frac{V_{cs}^\ast V_{cb}^R}{V_{ts}^\ast V_{tb}}\sin\zeta\cos\zeta 
                  e^{i\omega}\left(A_{\rm LR}(x_c)
                                -\frac{M_{W_1}^2}{M_{W_2}^2}A_{\rm LR}(\tilde{x}_c)
                          \right)
                 +\frac{m_t}{m_b}\frac{\tan(2\beta_{\rm LR})}{\cos(2\beta_{\rm LR})}
                  e^{i\omega}\frac{V_{tb}^R}{V_{tb}}h_7(y) \nonumber \\
             & & +\tan(2\beta_{\rm LR})A_H^2(x), \\
  C_7'{}^{\rm NP} &=& \frac{g_R^2}{g_L^2}\frac{V_{ts}^{R\ast}V_{tb}^R}{V_{ts}^\ast V_{tb}}
                   \left(\sin^2\zeta D_0'(x_t)+\cos^2\zeta
                         \frac{M_{W_1}^2}{M_{W_2}^2}D_0'(\tilde{x}_t)\right) \nonumber \\
                       & & +\frac{m_t}{m_b}\frac{g_R}{g_L}\frac{V_{ts}^{R\ast}}{V_{ts}^\ast}
                              \sin\zeta\cos\zeta e^{-i\omega}
                              \left(A_{\rm LR}(x_t)
                                     -\frac{M_{W_1}^2}{M_{W_2}^2}A_{\rm LR}(\tilde{x}_t)\right) \nonumber \\
                       & & +\frac{m_c}{m_b}\frac{g_R}{g_L}
                            \frac{V_{cs}^{R\ast}V_{cb}}{V_{ts}^\ast V_{tb}}\sin\zeta
                            \cos\zeta  e^{-i\omega}
                            \left(A_{\rm LR}(x_c)
                                  -\frac{M_{W_1}^2}{M_{W_2}^2}A_{\rm LR}(\tilde{x}_c)
                            \right) \nonumber \\
                       & & +\frac{m_t}{m_b}
                            \frac{\tan(2\beta_{\rm LR})}{\cos(2\beta_{\rm LR})}
                            e^{-i\omega}\frac{V_{ts}^{R\ast}}{V_{ts}^\ast}h_7(y)
                           +\frac{V_{ts}^{R\ast}V_{tb}^R}{V_{ts}^\ast V_{tb}}
                            \frac{1}{\cos^2(2\beta_{\rm LR})}A_H^2(x),
 \end{eqnarray}
when one does not assume $g_R/g_L=1$ and $V=V_R$, where loop functions $D_0'(x)$ 
and $A_{\rm LR}(x)$ are given in Appendix A, and $x_t\equiv m_t^2/m_{W_1}^2$, 
$\tilde{x}_t\equiv m_t^2/m_{W_2}^2$, $x_c\equiv m_c^2/m_{W_1}^2$, $\tilde{x}_c\equiv 
m_c^2/m_{W_2}^2$, and $y\equiv m_t^2/M_H^2$. 
The ratio $|C_7'/C_7|$ in the LRSM is 
shown by the blue curves in Fig.~\ref{fig3}. 
The value of the ratio in the LRSM is larger than the both cases of SM and SUSY 
SU(5) with $N_R$ model because the right-handed current in the LRSM is more effective than 
those models. In the analysis, we take 
 \begin{eqnarray}
  \frac{g_R}{g_L}=1,~~~V=V^R,~~~\tan\beta_{\rm LR}=10,~~~
  \omega=\frac{\pi}{10},~~~ M_{H^\pm} = 15 {\rm TeV},
 \end{eqnarray}
as a sample point.
\footnote{Although there are crossing points of the NP lines with the SM prediction line in all figures afterward, they are due to the fixing of the charged Higgs mass.}

\begin{figure}
\begin{center}
\includegraphics[scale=0.78]{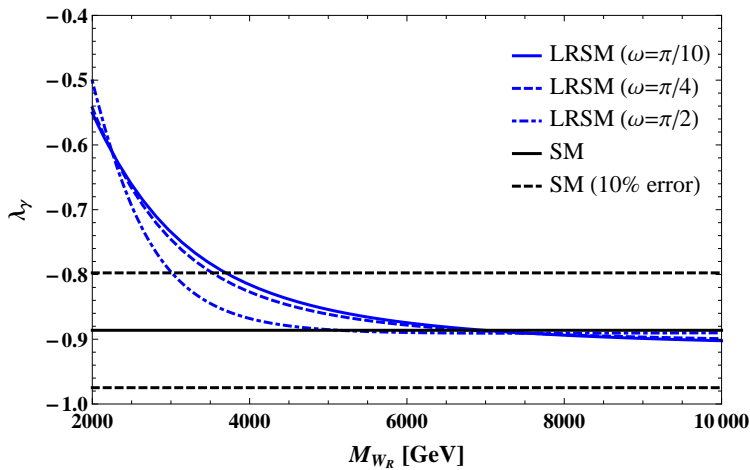}
\end{center}
\caption{The polarization parameter $\lambda_\gamma$ in $b\rightarrow s\gamma$ in
 the SM and LRSM which are depicted by the black and blue curves, respectively. 
We also show $\omega=\pi/4$ and $\pi/2$ cases by the blue dashed and 
dashed-dotted curves for comparison, respectively.}
\label{fig4}
\end{figure}

The polarization parameter in the LRSM is shown by the blue curve in 
Fig.~\ref{fig4}. 
The SM and LRSM cases are depicted by the black and blue curves, respectively. 
We also show $\omega=\pi/4$ and $\pi/2$ cases by the blue dashed and 
dashed-dotted curves for comparison, respectively. Since $M_{\rm 
NP}\simeq 3.7~{\rm TeV}$ for 10 $\%$ deviation from the SM prediction, the future
 LHCb experiment with 2 fb$^{-1}$ will check the scale of the right-handed gauge boson 
up to $3.7~{\rm TeV}$ in this case. The dependence of photon polarization on the CP phase is not strong.

\section{CP asymmetry}
Next, we evaluate the CP asymmetry in $b \to s \gamma$ process in each model. 
The CP asymmetry can be categorized into two parts: 
One is the direct CP asymmetry which is induced by the CP phase in the decay amplitude, 
and the other is the time-dependent CP asymmetry which is induced during the meson mixing.

\subsection{Direct CP asymmetry}

In addition to the determination of photon polarization, the observation of CP 
asymmetry in the $b\rightarrow s\gamma$ process is still sensitive to the 
existence of NP. Thus, we evaluate the CP asymmetry of the process in both 
models of SUSY SU(5) GUT with $N_R$ and LRSM. The asymmetry is given by 
 \begin{eqnarray}
  A_{\rm CP}(b\rightarrow s\gamma)
   &\equiv& \frac{\Gamma(B\rightarrow X_{\bar{s}}\gamma)
                  -\Gamma(\bar{B}\rightarrow X_s\gamma)}
                 {\Gamma(B\rightarrow X_{\bar{s}}\gamma)
                  +\Gamma(\bar{B}\rightarrow X_s\gamma)} \nonumber \\
   &\simeq& -\frac{1}{|C_7|^2+|C_7'|^2}
            (1.23~{\rm Im}[C_2C_7^\ast]-9.52~{\rm Im}[C_8C_7^\ast+C_8'C_7'{}^\ast]
             +0.10~{\rm Im}[C_2C_8^\ast]) \nonumber \\
   &      & -0.5~~~(\mbox{in}~\%), \label{acp}
 \end{eqnarray} 
which of course strongly depends on the CP-phases in the model where $C_2$ is 
the coefficient of the operator $O_2=(\bar{c}^\alpha \gamma_\mu P_Lb^\alpha)(\bar{c}^\beta\gamma^\mu 
P_Lb^\beta)$ in the effective Hamiltonian of $\Delta F=2$ transitions. 
Note that the contributions from NP to 
$C_2$ and $C_2'$ are negligibly small while $C_{7,8}$ and $C_{7,8}'$ include 
contributions from NP, i.e. $C_2'\ll 
C_2=C_2^{\rm SM}=1$. 
Thus, $A_{\rm CP}(b\rightarrow s\gamma)$ is well approximated by Eq.~(\ref{acp}).\footnote{There is also an error in the SM prediction but it is enough small~\cite{Hurth:2003dk}. Therefore, we neglect such correction here just for simplicity.}

\subsubsection{SUSY SU(5) with right-handed neutrino case}

\begin{figure}
\begin{center}
\includegraphics[scale=0.74]{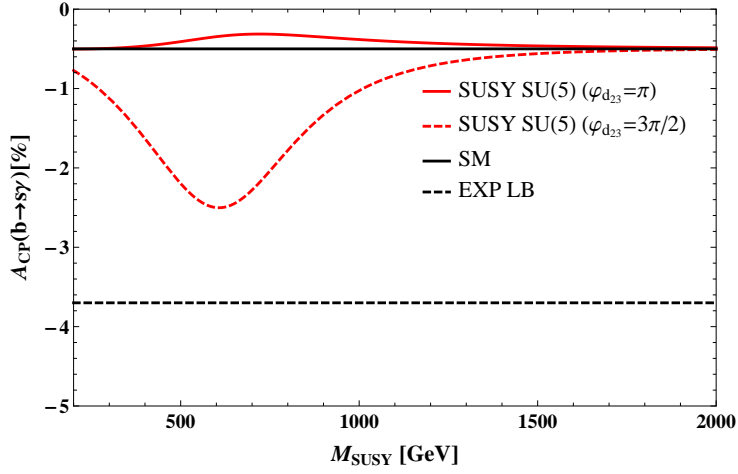}
\end{center}
\caption{The direct CP asymmetry in $b\rightarrow s\gamma$ in the SM and SUSY 
SU(5) GUT with $N_R$, which are depicted by the black and
 red solid curves. The black dashed line corresponds to the current experimental
 lower bound~\cite{Butler:2013kdw}. We also show the $\varphi_{d_{23}}=3\pi/2$ case by 
the red dashed curve for comparison.}
\label{fig5}
\end{figure}

We show $A_{\rm CP}(b\rightarrow s\gamma)$ in the SUSY SU(5) with $N_R$ and SM cases by red and black solid 
curves in Fig.~\ref{fig5}, respectively. 
The experimental lower bound as $-3.7~\%$~\cite{pdg} is also shown by the black 
dashed line. We also show $\varphi_{d_{23}}=3\pi/2$ case by the red dashed curve for 
comparison. It turns out that the magnitude of the direct CP asymmetry in the 
case of SUSY SU(5) with $N_R$ highly depend on the phase 
$\varphi_{d_{23}}$.
The current experimental lower bound does not constrain the scale of NP, $M_{\rm SUSY}$ 
in this parameter setup.
Therefore, the measurement of $A_{\rm CP}$ does not 
currently constrain on the NP scale in the SUSY SU(5) with $N_R$ model even if one takes 
the maximally allowed CP phase as $\varphi_{d_{23}}=3\pi/2$. But, the expected reach 
of Belle II with 50 ab$^{-1}$ for $A_{\rm CP}(B\rightarrow X_{s+d}\gamma)$ will be 
$\pm2\%$ precision. Thus, the future determination will check NP 
between $450 \mathchar`-  750$ GeV in this typical SUSY SU(5) with $N_R$ case.

\subsubsection{LRSM case}

For the LRSM, $C_8^{\rm NP}$ and $C_8'{}^{\rm NP}$ are 
\begin{eqnarray}
C_8 &=& \rho_8 \Delta^{LR}  C_8 
+ \rho_{\rm LR} \frac{m_c}{m_b} \sin \zeta \cos \zeta e^{i \alpha} \frac{V_{cb}^R}{V_{cb}^L}\,,\\
C_8' &=& \rho_8 \Delta^{LR}  C_8' 
+ \rho_{\rm LR} \frac{m_c}{m_b} \sin \zeta \cos \zeta e^{-i \alpha} \frac{V_{cb}^{R \ast}}{V_{cb}^{L \ast}}\,,
\end{eqnarray}
with
\begin{eqnarray}
\Delta^{LR}  C_8 &=& \frac{m_t}{m_b} \sin \zeta \cos \zeta e^{i \alpha} \frac{V_{tb}^R}{V_{tb}^L} f_{\rm LR} (\tilde{x}_t)\,,\\
\Delta^{LR} C_8' &=& \frac{m_t}{m_b} \sin \zeta \cos \zeta e^{-i \alpha} \frac{V_{ts}^{R \ast}}{V_{ts}^{L \ast}} f_{\rm LR} (\tilde{x}_t)\,,
\end{eqnarray}
where $\rho_8$ and $\rho_{\rm LR}$ are the so-called magic number which are given in the Ref.~\cite{Blanke:2011ry}. 
$f_{\rm LR} (x)$ is also a loop function for the left-right symmetric model given in Appendix A. 

\begin{figure}[t]
\begin{center}
\includegraphics[scale=0.78]{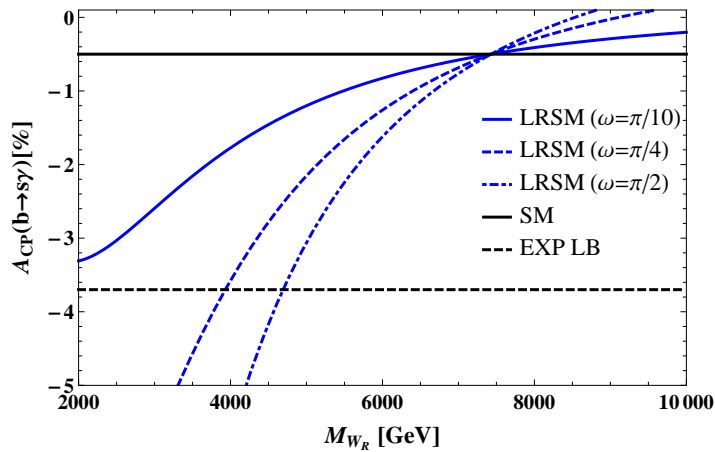}
\end{center}
\caption{The direct CP asymmetry in $b\rightarrow s\gamma$ in the SM and LRSM 
which are depicted by the black and blue solid curves. 
The black dashed line is same as in Fig.~\ref{fig5}. We also show $\omega=\pi/4$
 and $\pi/2$ cases by the blue dashed and dashed-dotted curves for comparison, 
respectively.}
\label{fig6}
\end{figure}

The asymmetry in the LRSM is shown by the blue curve in Fig.~\ref{fig6}. 
We also show $\omega=\pi/4$ and $\pi/2$ cases by the blue dashed and 
dashed-dotted curves for comparison, respectively. One can see from 
Fig.~\ref{fig6} that the magnitude of the direct CP asymmetry in the LRSM is 
allowed for $M_{\rm NP}\geq2~{\rm TeV}$ for $\pi/10$. 
Hence, the $A_{\rm CP}$ measurement does not give constraint on the existence of NP 
in this case at the moment. 
Furthermore, the future Belle II with  50 ab$^{-1}$ will check the LRSM model 
up to $3.5 {\rm TeV}$, $5 {\rm TeV}$, and $5.5 {\rm TeV}$ for 
$\omega = \pi/10$, $\omega = \pi/4$, and $\omega = \pi/2$, respectively. 
The result is really sensitive to the phase $\omega$ as with SUSY SU(5) with $N_R$ case.

\subsection{Time-dependent CP asymmetry}
We also evaluate the time-dependent CP asymmetry in 
the $B \to K_s \pi^0 \gamma$ decay denoted as $S_{CP}$.
The definition of $S_{CP}$ is same in the both model:
\begin{eqnarray}
S_{\rm CP} &=& 2 \frac{{\rm Im} \left[ e^{-2i \beta_{CKM}} C_7 C'_7 \right]}{\left|C_7\right|^2 + \left|C'_7\right|^2}, \label{scp}
\end{eqnarray}
where $2 \beta_{CKM} \approx 43^\circ$ is a CP phase in $B \to K_s \pi^0 \gamma$ decay.

\subsubsection{SUSY SU(5) with right-handed neutrino case}

\begin{figure}[t]
\begin{center}
\includegraphics[scale=0.8]{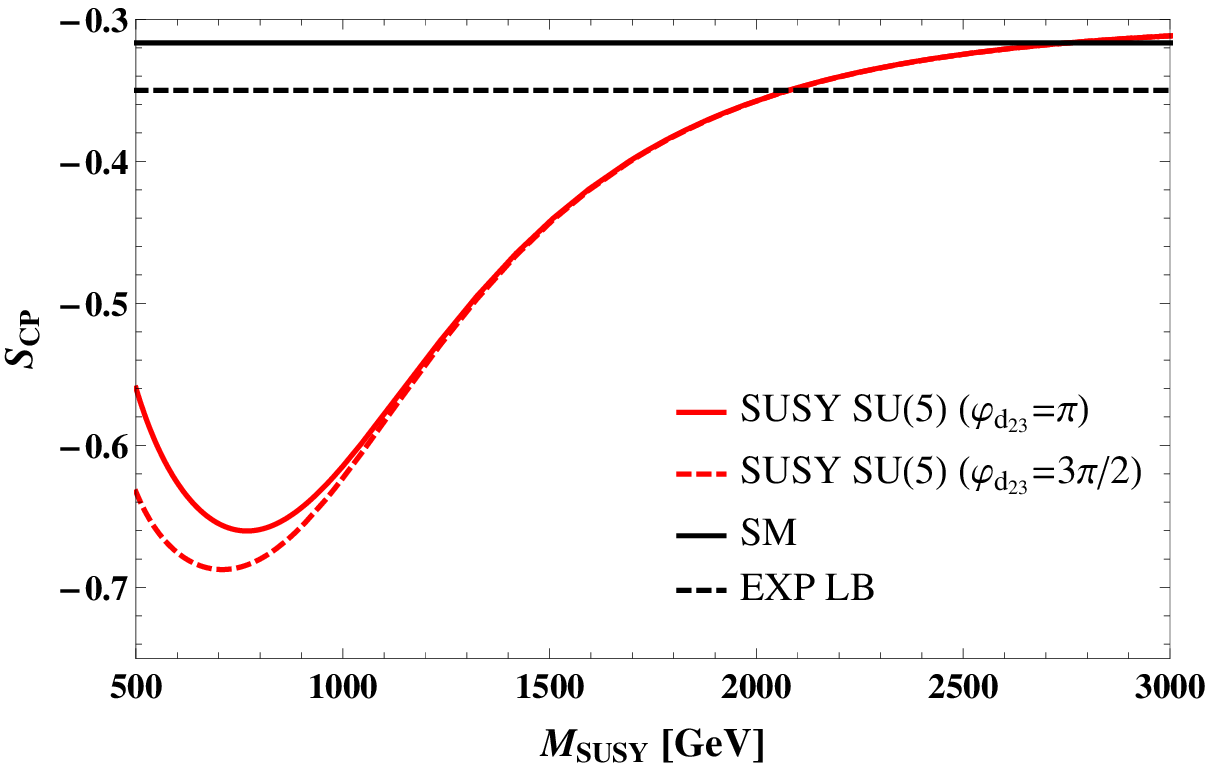}
\end{center}
\caption{The time-dependent CP asymmetry in $B \to K_s \pi^0 \gamma$ in the SM and SUSY SU(5) GUT with $N_R$, 
which are depicted by the black and blue solid curves. 
The black dashed line is the current experimental lower bound \cite{Butler:2013kdw}. We also show $\varphi_{d_{23}}=3\pi/2$ case by the red dashed curve for comparison.}
\label{fig7}
\end{figure}

We show Eq.~(\ref{scp}) in the SUSY SU(5) with $N_R$ and SM cases by red and black solid 
curves in Fig.~\ref{fig7}, respectively. 
The current experimental lower bound as $-0.35$ is also shown by the black dashed line.
We also show $\varphi_{d_{23}}=3\pi/2$ case by the red dashed curve for comparison.
The current experimental bound seems to exclude up to $2~ {\rm TeV}$ of the $M_{\rm SUSY}$ in Fig.~\ref{fig7} in this naive set up. 
In other words, the time-dependent CP asymmetry gives the strongest constraint 
within $\lambda_\gamma$, $A_{\rm CP}$, and $S_{\rm CP}$.

\subsubsection{LRSM case}

\begin{figure}[t]
\begin{center}
\includegraphics[scale=0.78]{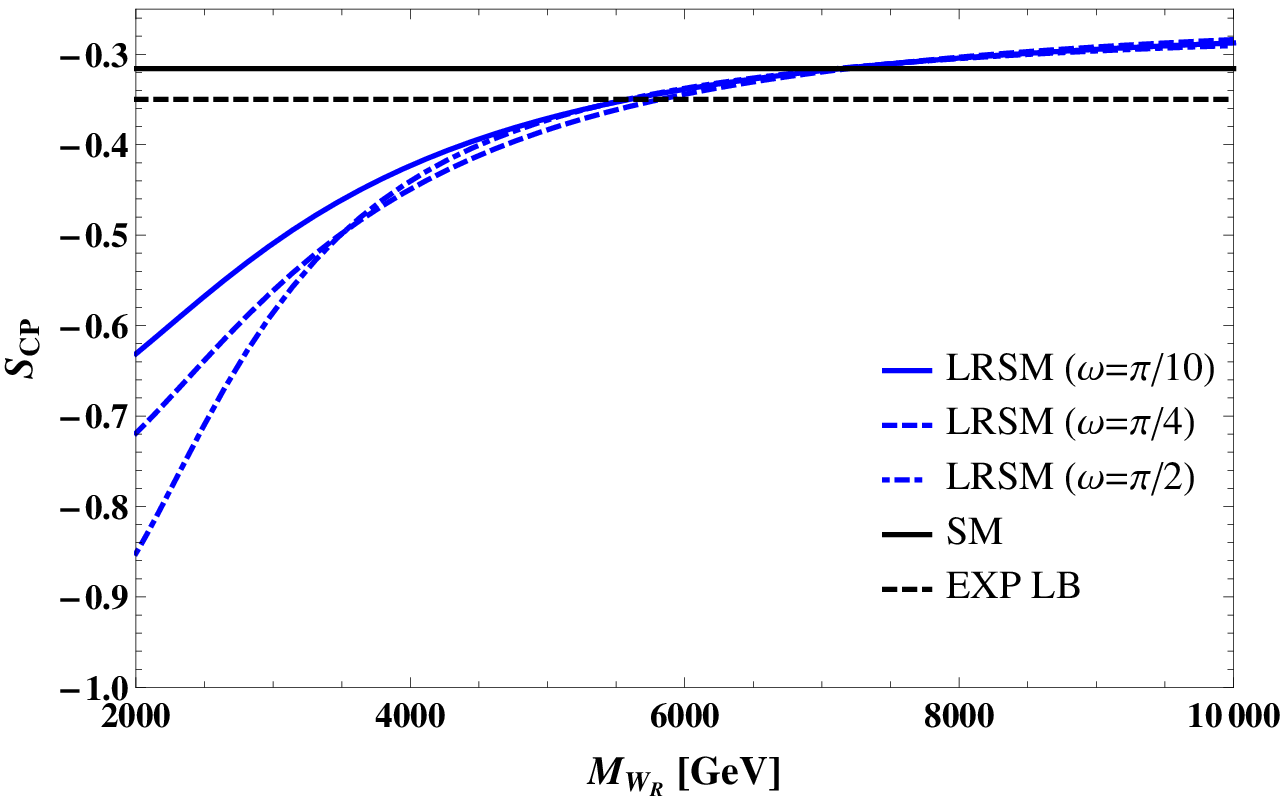}
\end{center}
\caption{The time-dependent CP asymmetry in $B \to K_s \pi^0 \gamma$ in the SM and LRSM 
which are depicted by the black and blue solid curves. 
The black dashed line is same as in Fig.~\ref{fig7}. We also show $\omega=\pi/4$
 and $\pi/2$ cases by the blue dashed and dashed-dotted curves for comparison, 
respectively.}
\label{fig8}
\end{figure}

We show Eq.~(\ref{scp}) in the LRSM and SM cases by blue and black solid 
curves in Fig.~\ref{fig8}, respectively. 
The current experimental lower bound constrains the mass of $W_R$ up to $7~{\rm TeV}$ 
in this simple set up.
The figure shows that the time-dependent CP asymmetry is the strongest constraint even in the LRSM.

\section{Comparison between Photon polarization and Measurements of CP asymmetry}

As we mentioned in the Introduction, 
the photon polarization might become a useful way to determine NP. 
Actually, the ways to determine the photon chirality by measurement of angular distribution 
of the final state particles have been discussed in several papers \cite{Melikhov:1998cd,Kim:2000dq,Becirevic:2012dx} (See also \cite{Li:2010ra}). 
In this paper, we have evaluated three observables: 
photon polarization, direct CP asymmetry, and time-dependent CP asymmetry.
We give some comments on the comparison among the results of those:
\begin{itemize}
\item For the determination of photon polarization in the SUSY SU(5) with $N_R$ model, 
the future LHCb experiment with $2~{\rm fb}^{-1}$ will be able to check the NP scale 
up to about 1700 GeV, which corresponds to $M_2\simeq 1400$ GeV and 
$M_{\tilde{g}} \simeq 4900$ GeV, at the typical point of this model.

\item For the measurement of $A_{\rm CP}$ in the SUSY SU(5) with $N_R$ model, 
the experiment does not currently constrain on the NP scale even if one takes the maximally 
allowed CP phase as $\varphi_{d_{23}}=3 \pi/2$. The future Belle II with 50 ab$^{-1}$ 
will check NP between $450 \mathchar`- 750$ GeV.

\item For the determination of photon polarization in the LRSM, the future LHCb 
experiment will check the NP scale up to $3.7~{\rm TeV}$ in this 
case.

\item For the measurement of $A_{\rm CP}$ in the LRSM, the experiment constrain 
up to $4.7~{\rm TeV}$ depending on the phase $\omega$. 
Furthermore, the future Belle II with 50 ${\rm ab}^{-1}$ will check up to $5.5~{\rm TeV}$ 
when $\omega$ maximize the direct CP asymmetry. 
On the other hand, the future experiment reach $3.5~{\rm TeV}$ at most 
when $\omega$ minimize the direct CP asymmetry. 

\item The time-dependent CP asymmetry is the most stringent constraint in both models 
and this CP asymmetry does not depend on CP asymmetry parameters so much.

\item Thus, we mention that there is a region where the determination of 
photon polarization is more ascendant for the NP search than that of direct CP 
asymmetry in both models. 
However, time-dependent CP asymmetry always gives us more stringent constraint 
than other observables. 
\end{itemize}

\section{Summary}

One might be able to obtain the existence of NP and discriminate among the SM 
and NP if one can precisely determine the chirality of the $s$ quark in the 
$b\rightarrow s\gamma$ process. The chirality of $s$ quark can be determined by 
measuring the polarization of photon in the process. 
And, there are several ways to measure the photon polarization. 
In addition to the determination of photon polarization, 
the observation of CP asymmetry in the process is still sensitive 
to the existence of NP. Thus, simultaneous studies of photon polarization and CP
 asymmetry in the $b\rightarrow s\gamma$ process will be intriguing for the 
experimental search of NP.

We have investigated the $b\rightarrow s\gamma$ process in the SUSY SU(5) GUT 
with $N_R$ model and the LRSM. The ratio $|C_7'/C_7|$, the 
polarization parameter of photon, and the direct CP asymmetry in the process 
have been evaluated in both models. 
The time-dependent CP asymmetry seems to be the best way to find the NP effect. However, the combination of CP violation and photon polarization can discriminate NP beyond the SM at the end.
Furthermore, there might be a region where the determination of photon polarization is
 more sensitive for the new physics search than that of CP asymmetry.

\subsection*{Acknowledgement}

We would like to thank to Y. Okada for fruitful discussion.
This work is partially supported by Scientific Grant by Ministry of Education 
and Science, No. 24540272 and Grant-in-Aid for Scientific Research on Innovative Areas titled ``Unification and　Development of the Neutrino Science Frontier'', No. 25105001. 
The work of R.T. are supported by Research 
Fellowships of the Japan Society for the Promotion of Science for Young 
Scientists.

\appendix
\section*{Appendix}

\section{Loop functions}

We give loop functions \cite{Altmannshofer:2009ne} which are utilized in our 
analyses:
 \begin{eqnarray}
  h_7(x)          &=& -\frac{5x^2-3x}{12(1-x)^2}-\frac{3x^2-2x}{6(1-x)^3}\log x, 
                      \\
  h_8(x)          &=& -\frac{x^2-3x}{4(1-x)^2}+\frac{x}{2(1-x)^3}\log x, \\
  f_{7,8}^{(1)}(x,y) &=& \frac{2}{x-y}(f_{7,8}^{(2)}(x)-f_{7,8}^{(2)}(y)), \\
  f_7^{(2)}(x)     &=& -\frac{13-7x}{24(1-x)^3}-\frac{3+2x-2x^2}{12(1-x)^4}\log x,
                     \\
  f_8^{(2)}(x)     &=& \frac{1+5x}{8(1-x)^3}+\frac{x(2+x)}{4(1-x)^4}\log x, \\
  f_{\rm LR}(x) &=&  - \frac{x^2 + x + 4}{4 (1-x)^2} - \frac{3x}{2(1-x)^2} \log x, \\
  g_7^{(1)}(x)     &=& -\frac{2(1+5x)}{9(1-x)^3}-\frac{4x(1+x)}{3(1-x)^5}\log x, 
                     \\
  g_8^{(1)}(x)     &=& \frac{11+x}{3(1-x)^3}+\frac{9+16x-x^2}{6(1-x)^4}\log x, \\
  g_7^{(2)}(x)     &=& -\frac{2(1+10x+x^2)}{9(1-x)^4}
                     -\frac{4x(1+x)}{3(1-x)^5}\log x, \\
  g_8^{(2)}(x)     &=& \frac{53+44x-x^2}{12(1-x)^4}
                     +\frac{3+11x+2x^2}{2(1-x)^5}\log x, \\
  D_0'(x)         &=& \frac{-8x^3-5x^2+7x}{24(x-1)^3}
                     +\frac{3x^3-2x^2}{4(x-1)^4}\log x, \\
  A_{\rm LR}(x)     &=& \frac{-5x^2+31x-20}{6(x-1)^2}
                      -\frac{3x^2-2x}{(x-1)^3}\log x, \\
  A_H^2(x)        &=& \frac{22x^3-53x^2+25x}{72(x-1)^3}
                      -\frac{3x^3-8x^2+4x}{12(x-1)}\log x.
 \end{eqnarray}

\section{Mass insertion parameters}

The mass insertion parameters are defined as (e.g., see~\cite{Hisano:2004pw})
 \begin{eqnarray}
  (\delta_q^{XX})_{ij} &\equiv& \frac{(m_{\tilde{q}_{X}}^2)_{ij}}{\tilde{m}^2_{\tilde{f}}}, \label{Eq:mass_ins_same}\\
  (\delta_d^{XY})_{ij} &\equiv& \frac{v_d(A_d-\mu t_\beta)_{ij}}{\tilde{m}^2_{\tilde{f}}}, \\
  (\delta_u^{XY})_{ij} &\equiv& \frac{v_u(A_u-\mu\cot)_{ij}}{\tilde{m}^2_{\tilde{f}}},
 \end{eqnarray}
with $X\neq Y$ where $\tilde{m}^2_{\tilde{f}}\, ({\tilde{f}}=\tilde{u}\,,\tilde{d})$ denotes up- and down-type averaged squark mass, 
and the numerator of Eq. (\ref{Eq:mass_ins_same}) can be written as
 \begin{eqnarray}
  (m_{\tilde{u}_L}^2)_{ij} 
  &\simeq& -V_{3i}V_{3j}^\ast\frac{f_t^2}{(4\pi)^2}(3m_0^2+A_0^2)
           \left(2\log\frac{M_{\rm pl}^2}{M_{H_c}^2}
                 +\log\frac{M_{H_c}^2}{M_{\rm SUSY}^2}\right), \\
  (m_{\tilde{u}_R}^2)_{ij}  
  &\simeq& -e^{-i\varphi_{u_{ij}}}V_{3i}^\ast V_{3j}\frac{2f_b^2}{(4\pi)^2}(3m_0^2+A_0^2)
           \log\frac{M_{\rm pl}^2}{M_{H_c}^2}, \\
  (m_{\tilde{d}_L}^2)_{ij}  
  &\simeq& -V_{i3}^\ast V_{j3}\frac{2f_t^2}{(4\pi)^2}(3m_0^2+A_0^2)
           \left(3\log\frac{M_{\rm pl}^2}{M_{H_c}^2}
                 +\log\frac{M_{H_c}^2}{M_{\rm SUSY}^2}\right), \\
  (m_{\tilde{d}_R}^2)_{ij}  
  &\simeq& -e^{-i\varphi_{d_{ij}}}U_{ki}^\ast V_{kj}\frac{f_{\nu_k}^2}{(4\pi)^2}
           (3m_0^2+A_0^2)\log\frac{M_{\rm pl}^2}{M_{H_c}^2}.
 \end{eqnarray}
Here, $m_0$ and $A_0$ are the universal scalar mass and trilinear coupling, 
respectively.

\end{document}